\documentclass[notitlepage,prd,superscriptaddress,nofootinbib]{revtex4-2}
\usepackage{amsmath} 
\usepackage{xcolor}
\usepackage{bbold}
\def\bphi{\bar{\phi}}

\def\be{\begin{equation}}
\def\ee{\end{equation}}
\def\arr{\begin{array}{rll}}
\def\ea{\end{array}}
\def\bea{\begin{eqnarray}}
\def\eea{\end{eqnarray}}
\def\cN{{\cal{N}}}

\def\N2{$N{=}2$}

\def\>{\rangle}
\def\<{\langle}
\def\+{\dagger}
\def\={\ =\ }

\begin{document}
\renewcommand{\thefootnote}{\arabic{footnote}}
\setcounter{footnote}0
\setcounter{page}{1}
\title{ Note on   $\mathcal{N}=8$ supersymmetric mechanics with dynamical and semi-dynamical multiplets}
 \author{Erik Khastyan }
 \email{khastyan.erik@gmail.com}
 \affiliation{Yerevan Physics Institute, 
2 Alikhanyan Brothers  St., Yerevan, 0036, Armenia}
\author{Sergey Krivonos}
\email{krivonos@theor.jinr.ru}
\affiliation{Bogoliubov Laboratory of Theoretical Physics, Joint Institute for Nuclear Research,
Dubna, Russia}
\author{Armen Nersessian}
\email{arnerses@yerphi.am}
\affiliation{Yerevan Physics Institute,
2 Alikhanyan Brothers  St., Yerevan, 0036, Armenia}
\affiliation{Bogoliubov Laboratory of Theoretical Physics, Joint Institute for Nuclear Research,
Dubna, Russia}
\affiliation{Institute of Radiophysics and Electronics, Ashtarak-2, 0203, Armenia }
\begin{abstract}
 We give a Hamiltonian formulation of the new model of  $\mathcal{N}=8$ supersymmetric mechanics recently proposed by  S.~Fedoruk and E.~Ivanov \cite{IvFed} and show that it  possesses  the dynamical $\mathcal{N}=8$ superconformal symmetry $osp(8|2)$. The bosonic part of the Hamiltonian is just  a free particle on   eight-dimensional cone embedded in nine-dimensional pseudo-Euclidean space, while the fermionic part  can be interpreted  as a  spin-orbit interaction term.

\end{abstract}
\maketitle

%

\section{Introduction}
 
The natural framework for constructing and dealing with supersymmetric mechanics is the  superfield approach. It works  perfectly in the case of lower,  $\cN \leq 4$ supersymmetries, but   acquires a lot of complications for theories with $\cN=8$ supersymmetry. The reason is simple - the measure of the corresponding $\cN=8$ superspace is too large  to be used to write   superfield actions. To resolve this problem, the evident way is to construct the action in terms of several $\cN=4$ superfields so as  to have another implicit $\cN=4$ supersymmetry which together with the explicit supersymmetry would  form $\cN=8$ supersymmetry [1-9]. Clearly, such a construction of  $\cN=8$  supersymmetric systems is much more complicated than $\cN=4$ supersymmetric ones because
the number and content of $\cN=4$ superfields involved is unknown a priori. Moreover, in some cases to have an appropriate component action, one has to use the so called oxidation procedure to avoid the second-order Lagrangian for fermions. Thus, the final component action, with excluded auxiliary
fields, is related to  the superfield Lagrangian in an  indirect and highly non-trivial way. The natural way to clarify the properties of the final component action and its possible (hidden) symmetries is to construct a  Hamiltonian formulation and perform a proper redefinition of the fields involved. 

Another approach to the $\cN=8$ supersymmetric mechanics has been developed in \cite{toppan1}, where
the most general $\cN=8$ invariant action was constructed based on the   $\mathbf{(1,8,7)}$ supermultiplet.  It depends on two real parameters: $a,b$ and gives an interaction term; by setting $a=0$ one recovers the action of ref. \cite{ABC}.  Later on it was shown in \cite{toppan2} 
that the $F(4)$ superconformal action of Delduc-Ivanov \cite{F4scm} is recovered
by setting $b=0$ in the action in  \cite{toppan1}. Finally, in 2018, the quantum superconformal Hamiltonian of the $F(4)$ model was constructed in \cite{toppan3}.

Recently, S.~Fedoruk and E.~Ivanov have presented the new model of $\mathcal{N}=8$ supersymmetric mechanics constructed by using  the so-called   dynamical and semi-dynamical $\mathcal{N}=4$ multiplets \cite{IvFed}. The final multiplet content of the  model is $\mathbf{(1,8,7)}\bigoplus \mathbf{(8,8,0)}$. 
After eliminating  auxiliary variables,  the authors arrived the Lagrangian depending on the single real bosonic variable, the eight complex bosonic variables, and  the eight  real fermionic variables. Only the real bosonic variable   appears in the Lagrangian  with the second-order kinetic term, while the kinetic terms corresponding to  complex  bosonic variables are  of the first order.
Hence,   the system   could be  directly  formulated on the $(18|8)_{\mathbb{R}}$ dimensional phase space.
The authors interpreted the model as a $\mathcal{N}=8$ supersymmetric extension of the  one-dimensional isospin particle. 

In this paper,  we  give the Hamiltonian formulation of this model
and find two sets of  constants of motion: the odd constants of motion - supercharges forming  the $d=1,\mathcal{N}=8$ Poincare superalgebra and the  even constants of motion forming  $so(8)$ algebra. The bosonic and fermionic parts of these  $so(8)$-generators   separately define the constants of motion and form the same $so(8)$ algebra.
We also demonstrate that   the  supercharges   have the structure
``$so(8)-generators\, \times\, fermions$''.  

Moreover, we indicate two unexpected properties of the model:
\begin{itemize}
\item It possesses dynamical  $osp(8|2)$ superconformal symmetry.
\item
It describes the specific $\mathcal{N}=8$ supersymmetric extension of the free particle on {\sl eight-dimensional cone } embedded in nine-dimensional pseudo-Euclidean space, with the
 fermionic part, which can be interpreted as a spin-orbit coupling term.
 \end{itemize}
Thus, the proposed model is distinguished  among the conventional  supersymmetric mechanics models, where   supersymmetry arises due to the Pauli term  defining the interaction of spin with external   fields.\\

The paper is organized as follows:

In {\sl Section 2}, we give the Hamiltonian formulation of the model,
construct its supercharges and  even constants of motion. Then we  show that this variant of $\cN=8$ supersymmetric mechanics  possesses   $so(8|2)$ dynamical symmetry.

In {\sl Section 3}, we show that the bosonic part of the  model is a  free particle on eight-dimensional cone embedded in nine-dimensional pseudo-Euclidean space, while the fermionic part  can be interpreted as a spin-orbital interaction term.

In {\sl Conclusion } we review obtained results and mention possible future developments

In  the {\sl Appendix  A}   we present the explicit expressions for the superconformal algebra $osp(8|2)$ written in the notation used.

\section{Hamiltonian formulation}
\noindent
The model proposed in \cite{IvFed} after eliminating   auxiliary variables, 
is defined by the following  Lagrangian \footnote{For   convenience we have divided  by two  the Lagrangian given by Eq.4.30 in \cite{IvFed} and re-denoted $x$ by $r$. 
 Besides, we simplified it and used a  slightly different notation.} \cite{IvFed}:
\be 
L= \frac{\dot{ r}^2}{2}   + \frac{  r}{2} 
\left(z^{aiA}\dot z_{aiA}+ \dot B^{i' a} \bar B_{i' a} -B^{i' a} \dot{\bar B}_{i' a}\right)
-\frac{\imath}{2}\chi^{i' A}\dot\chi_{i' A}-\frac{\imath}{2} {\bar\phi}^{i}\dot\phi_{i}+ \frac{\imath}{2} \dot{\bar\phi}^{i}\phi_{i}-   I(B,\bar B, z)- \frac{1}{ r}{ {\mathcal{F}(\phi,\bar\phi, \chi,B,\bar B, z)}}
\label{L-comp}\ee
with
\be
 {I}  = \frac{1}{8} \left(B^{i' a}B_{i' a} \bar B^{j' b}\bar B_{j' b} -\left(B^{i' a}\bar B_{i' a}\right)^2 +
z_{akB}z_b^{kB}B^{m' a}\bar B_{m'}^b+ \frac{1}{2} z_{aiA}z_b^{iA}z^a_{jB}z^{bjB} \right)  \label{IB}
\ee
and 
\be
\mathcal{F} =\frac{\imath}{4} \left[ 
 (\bar\phi^{i}B^{k' a}- \phi^{i}\bar B^{k' a}+\chi^{k' A}z^{a i}_{A} )
 (\bar\phi_{i}B_{k' a}- \phi_{i}\bar B_{k' a}+\chi^{B}_{k'} z_{aiB} )
-2\left(\chi^{i' A}B_{i'}^a- \phi_i z^{aiA}\right) 
\left(\chi^{k'}_{A}\bar B_{a k'}+\bar\phi^k z_{akA}\right)\right],
\label{Fk}
\ee
where  $r$ is real bosonic coordinate, $z^{iA}_a, B^{i'}_a$ are the  complex  bosonic coordinates and $\chi^{i' A},\phi^i$ are fermionic ones; 
$i,j,k,i'=1,2$, $A,B=1,2$ are the spinorial indices, which
can be raised and lowered by the spinorial metric $\epsilon_{ij}$ and its inverse $\epsilon^{ij}$
$
$.
Besides, the indices  $a,b=1,2$, which number the fields involved, are summarized with the help of the "spinorial" 
 $\epsilon_{ab}$. Thus, all  expressions of the form $ \epsilon_{ab} A_a B_b$  can be rewritten as
  $ \epsilon_{ab} A_a B_b = \epsilon^{ba} A_a B_b = A^a B_a $.
The notation for conjugation properties  of the basic variables is as follows:
\be
\begin{array}{c} \left( B^{i'}_a \right)^\dagger ={\bar B}_{i' a}, \quad \left( B_{i' a} \right)^\dagger =-{\bar B}^{i'}_a, \quad 
 \left( z^{i A}_a \right)^\dagger = z_{a i A}, \quad  \left( z^{i}_{a A} \right)^\dagger = -z^A_{a i}, \quad  \left( z^{A}_{ a i} \right)^\dagger = -z^i_{a A}\nonumber \\
  \left( \phi^i\right)^\dagger = \bphi_i, \quad \left( \phi_i\right)^\dagger = -\bphi^ i, \quad \left( \chi^{i' A}\right)^\dagger = \chi_{i' A}, \quad\left(\chi^{i'}_A\right)^\dagger = -\chi_{i'}^A, \quad \left( \chi_{i'}^A\right)^\dagger = -\chi^{i'}_A.
\end{array}\ee

For the Hamiltonian formulation of the model we at first   re-scale the bosonic  variables  
\be
\begin{array}{cc}
 \sqrt{ {|r|} }  z^{iA}_a\quad\to\quad z^{iA}_a,\qquad \sqrt{ {|r|}  }  B^{i'}_a\quad\to\quad {B_{a}^{i'}} ,&\qquad {\rm when}\quad r>0\\
 \sqrt{ {|r|} }  z^{iA}_a\quad\to\quad (-1)^{a} z^{iA}_a,\qquad \sqrt{ {|r|}  }  B^{i'}_a\quad\to\quad (-1)^{a}{B_{a}^{i'}} ,&\qquad  {\rm when}\quad r<0
\end{array} .\ee 
The we  replace the first term in the Lagrangian by the variationally equivalent one
\be
\frac{\dot{r}^2}{2}\quad\to\quad p_r\dot{r}-\frac{p^2_r}{2}.
\ee
As a result, we   get the first-order Lagrangian 
\be
L=\mathcal{A}_{I}\dot{y}^I -\mathcal{H}, 
\ee
where   
\be
\mathcal{A}_{I}d{y}^I:=p_rd{r}+\frac{ 1}{2}
\left(z^{aiA}d z_{aiA}+ \bar B_{i' a}dB^{i' a} -B^{i' a} d{\bar B}_{i' a}\right)
-\frac{\imath}{2}\chi^{i' A}d\chi_{i' A}+\frac{\imath (\phi_{i}d{\bar\phi}^{i}-{\bar\phi}^{i}d\phi_{i})}{2}
\ee
and  
\be
\mathcal{H}=\frac{p^2_r}{2}+\frac{\mathcal{I}}{r^2},\qquad \mathcal{I}=I(B,\bar B, z)+ \mathcal{F}(\phi,\bar\phi, \chi,B,\bar B, z) 
\label{Hm}\ee
with $I$ and $\mathcal{F}$   defined in \eqref{IB} and \eqref{Fk}.

It is seen that after such reformulation the Lagrangian becomes independent on  ${\rm sgn}\;r$.

 {\sl Hence,  in our further consideration we will     assume, without lost of generality, that $r>0$}.\\

From the above  first-order  Lagrangian we  immediately arrive  at  the Hamiltonian formulation of the system: it is given   by the Hamiltonian \eqref{Hm} and the canonical symplectic structure
\be
\Omega=d\mathcal{A}= dp_r\wedge d{r}+   \frac 12 dz ^{aiA}\wedge d z_{aiA}+    d\bar B_{i' a}\wedge dB^{i' a} 
-\frac{\imath}{2}d\chi^{i' A}\wedge d\chi_{i' A}+  \imath d\phi_{i}\wedge d{\bar\phi}^{i} . 
\label{ss}\ee
The respective   Poisson brackets   are defined  by the relations
\be 
  \{ p,r \} =1, 
\qquad  \{ B^{i' a}, {\bar B}_{j' b}  \} = \delta^{i'}_{j'} \delta^a_b, \qquad  \{ z^{aiA}, z_{bjB} \} =
2\delta^a_{b} \delta^i_j \delta^A_B,\qquad  \{ \phi^i, \bar{\phi}_j  \} = \imath \, \delta_j^i, \qquad \{ \chi^{i' A}, \chi_{j' B} \}= 2\,\imath \, \label{pb0}\delta_{j'}^{i'} \delta_B^A\;  .
\ee
We can then easily construct the supercharges that provide the system 
with  $\mathcal{N}=8$ Poincar\'e supersymmetry 
\be
\left\{Q^i , {\bar Q}^j \right\} = -2 \imath\epsilon^{ij} {\cal H}, \quad   \left\{Q^{i' A}, Q^{j' B}\right\} = 4   \imath \epsilon^{i'j'} \epsilon^{AB} \, {\cal H}.
\ee
\be
 Q^i =p_r\phi^i-\frac{\Theta^i}{r},\qquad {\bar Q}_i=p_r\bar\phi_i+\frac{ \overline{\Theta}_i}{r}\;,\qquad  Q^{i' A}=p_r\chi^{i' A}+\frac{ \Theta^{i' A}}{r}
\label{sc}
\ee
where
\bea
\Theta^i&=& \frac{1}{2} \left[ \phi^i B^{k'a} {\bar B}_{k' a}-    
  \frac{1}{2}   \phi^j z_a^{iA} z^a_{jA}-  \bar\phi^i B^{k' a} B_{k' a}-\chi^{j' A} z_{aA}^i B_{j'}^a\right],
  \nonumber\\ 
\overline{\Theta}^i&=&\frac{1}{2} \left[{\bar\phi}^i {\bar B}^{k' a}B_{k'a}  -\frac12\bar\phi_j z^{ai}_{A} z_a^{jA}-\phi^i\bar B_{k'a}\bar  B^{k'a}-\chi_{j' A} z^{a i A} \bar B^{j'}_a\right],\label{Theta}\\ 
\Theta^{i' A}&=&  \frac{1}{2} \left[ \chi_{k'}^A\left( B^{i' a}\bar B^{k'}_a+ B^{k' a} \bar B^{i'}_{a}\right)+\frac12\chi^{i' B} z_a^{j A} z^a_{j B}+ z_{a j}^A\left(\bar\phi^j B^{i' a} -\phi^j \bar B^{i' a}\right) \right].\nonumber
\eea
Looking at the Hamiltonian \eqref{Hm}  we can immediately conclude 
that the system possesses the   dynamical  conformal symmetry   \eqref{ca}
given by the generators (see, e.g. \cite{conformal})
\be 
\mathcal{H}=\frac{p^2_r}{2}+ \frac{\mathcal{I}}{r^2},\quad D=\frac{p_r r}{2},\quad K=\frac{r^2}{2}.
\label{hdk}\ee
To extend this algebra to the superconformal one, we introduce the conformal supercharges \cite{superconformal}
\be
  S^i = r\phi^i,\qquad {\bar S}_i = r \bar\phi_i, \qquad S^{i A} = r \chi^{i A}.
\label{cch}
\ee
However, these  generators do not form a closed algebra with \eqref{hdk} and \eqref{cch}.
To close the  superalgebra  of generators \eqref{sc},\eqref{cch} and \eqref{hdk},  we are forced to introduce   additional bosonic generators  
\footnote{These generators  obey the following conjugation rules:
$$ \left( J\right)^\dagger = {\bar J}, \quad\left( {\bar J}\right)^\dagger = J, \quad
\left( J_3\right)^\dagger = {J_3}, \quad 
\left( W^{ij}\right)^\dagger = W_{ij}, \; \left( X^{i'j'}\right)^\dagger = X_{i'j'}, \; 
\left( Y^{AB}\right)^\dagger = Y_{AB}, \quad \left( V^{i j' A}\right)^\dagger = {\bar V}_{i j' A}, 
$$}
\bea 
 && J = \frac{1}{2} B^{i' a} B_{i' a} + \frac{\imath}{2} \phi^i \phi_i, \quad {\bar J} = \frac{1}{2} \bar B^{i' a} \bar B_{i' a} + \frac{\imath}{2} \bar\phi^i \bar\phi_i,\quad J_3= - \frac{\imath}{2}   B^{i' a} \bar B_{i' a} + \frac{1}{2}\phi^i \bar\phi_i, \label{l1}\\
&& X^{i'j'} = \frac{1}{2}\left( B^{i' a} \bar B^{j'}_a +B^{j' a} \bar B^{i'}_a\right) +\frac{\imath}{4} \chi^{i' A} \chi^{j'}_A, \label{l2} \\
&&W^{ij} = \frac{1}{4} z^{a i A} z_{aA}^j+\frac{\imath}{2}\left( \phi^i\bar\phi^j + \phi^j\bar\phi^i \right)
,  \label{l3}\\
&& Y^{AB} =\frac{1}{4} z^{a i A} z_{ai}^B +\frac{\imath}{4} \, \chi^{k' A} \chi_{k'}^B,   \label{l4}\\
&& V^{i j' A} = z_a^{i A} B^{j' a }- \imath \phi^i \chi^{j' A} , \quad \;{\bar V}^{i j' A} = z_a^{i A} \bar B^{j' a }- \imath \bar\phi^i \chi^{j' A}. \label{l5}
\eea
Each set of the  generators presented on  the lines \eqref{l1}-\eqref{l4}  forms the $su(2)$ algebra,  while their mutual Poisson brackets vanish.  In contrast with them, the 
generators  \eqref{l5} transform under fundamental representations of these $su(2)$ groups.
Altogether  they form the $so(8)$ algebra of $R$-symmetry  given by   relations \eqref{so81}-\eqref{so8last}.

In the above terms, the  Casimir  of the  $so(8)$ algebra reads
\be\label{Cas}
 {\cal C} =  2 J {\bar J} + 2 J_3 J_3 + X^{i'j'} X_{i'j'}+ W^{ij}W_{ij} +  Y^{AB}Y_{AB} + V^{ijA} {\bar V}_{ijA}.
\ee
Note that it does not contain the fourth powers of fermions.

The angular part of the Hamiltonian \eqref{Hm} is expressed  via the $so(8)$ Casimirs in the following way:
\be 
\mathcal{ I} =  \frac{1}{8 } \left(5{\cal C}_{bos} - {4} {\cal C}\right),\qquad {\rm where }
\qquad 
{\cal C}_{bos} =\left. {\cal C}\right|_{ {\phi_i, \bphi_i, \chi_{i,A}}\to\;0}.
\label{Ii}\ee
 It is obvious that the bosonic and fermionic parts of the $so(8)$ generators \eqref{l1}-\eqref{l5}  commute with each other and 
  each of them  forms  the $so(8)$ algebra. They commute 
 with the bosonic part of above the Casimir,  ${\cal C}_{bos} =\left. {\cal C}\right|_{ {\phi_i, \bphi_i, \chi_{i,A}}\to\; 0}$, and with its fermionic part as well.
Therefore, the   $so(8)$ generators \eqref{l1})-\eqref{l5}  commute with the   generators of the conformal algebra \eqref{hdk} (and with the "angular Hamiltonian"  $\mathcal{I}$ ) and  define the  constants of motion of the system. 

The  above $so(8)$ generators together with the  generators of conformal algebra  \eqref{hdk}  and   supercharges \eqref{cch},\eqref{sc} form   the conformal $osp(8|2)$ superalgebra. 

 Thought the    superconformal symmetry  was not observed in \cite{IvFed}, this  is not surprising since the model under consideration was suggested as an $\mathcal{N}=8$ extension of the $\mathcal{N}=4$ superconformal mechanics suggested in \cite{FIL-2009}.

Let us complete this Section by expressing 
the angular parts of the supercharges \eqref{Theta}   via  $so(8)$ generators    in a form similar to those in   \cite{superconformal}
\be  
\begin{array}{c}
 \Theta^i = \imath J_3\phi^i -   W^{ij}\phi_j - J\bar\phi^i -\frac{1}{2}V^{ij'A}\chi_{j' A} , \quad
  {\bar\Theta}^i = \imath J_3\bar\phi^i  +  W^{ij}\bar \phi_j -  {\bar J}\phi^i +\frac{1}{2}{\bar V}^{ij'A}\chi_{j' A} , \\[3mm]
 \Theta^{i' A} =   \chi_{j'}^A X^{i'j'}+
\chi^{i'}_B Y^{AB} + \phi_j {\bar V}^{j i' A}-\bar\phi_j V^{j i' A} . 
\end{array}\label{Qadv}
\ee
In the next Section we   discuss  some consequences of the presented observations.

\section{Geometry and dynamics}
Since the system under consideration has   the canonical symplectic structure \eqref{ss}, 
  we can  interpret  the pairs $(z_1^{iA}, z_{2\;iA})$ and $ (B^i_1, \bar B_{2\; i})$
 as  canonically conjugated momenta and coordinates.
 For  visualization of this observation (and of the structure of the respective Hamiltonian system) we  suggest the following notation:
\begin{equation}
  \begin{array}{c}
		 z_1^{iA}: =x^4\delta^{iA} + \imath x^\alpha(\sigma_\alpha)^{iA}, \quad 
        z_{2\; iA} := p_4\delta_{iA} + \imath p_\alpha (\sigma_\alpha)_{iA} ,  \quad\chi^{iA}:=  \chi^4\delta^{iA} + \imath \chi^\alpha(\sigma^\alpha)^{iA} ,\qquad \alpha=1,2,3
		\\[3mm]
 B_1^{i} :=\frac{1}{\sqrt{2}}(x^{2i+3}+\imath x^{2i+4}),   \quad
        \bar B_{2\;i} := \frac{1}{\sqrt{2}} (p_{2i+3}-\imath p_{2i+4}),\quad     \phi_i:=\frac{1}{\sqrt{2}}(\chi^{2i+3}+\imath \chi^{2i+4}),\qquad i=1,2.
	\end{array}
\label{piz}\end{equation}
In these terms,   the Poisson brackets \eqref{pb0}  read
\be
\{p_r,r\}=1,\qquad \{p_\mu,x^\nu\}=\delta_\mu^\nu,\qquad  \{\chi^{\mu} ,  {\chi}^\nu\}=\imath\delta^{\mu\nu},
\ee
 while the bosonic and fermionic parts of $so(8)$ generators \eqref{l1}-\eqref{l5}  result, respectively in $L_{\mu\nu}:= p_\mu x_\nu-p_\nu x_\mu $, and $  R_{\mu\nu}:=\imath \chi_\mu\chi_\nu$.
The bosonic part of  \eqref{Ii} acquires now the form 
\be 
\mathcal{ I}_{bos}=\left. {\cal I}\right|_{\chi^\mu \to\;0} =\frac{1}{16} \sum_{\mu,\nu=1}^8  L^2_{\mu\nu},  
.
\label{Ii1}\ee
Here, the indices $\mu,\nu$  in $L_{\mu\nu}$  are raised/lowered by the use of Euclidian metrics $\delta_{\mu\nu}$. 
Being expressed in terms of  spherical coordinates and conjugated momenta the angular part 
becomes  independent of the radius $R=\sqrt{\sum_{\mu=1}^8 {x}^\mu {x}^\mu} $ and conjugated momentum $p_R= {\sum_{\mu=1}^8 {p}_\mu {x}^\mu}/R$.

Performing the  canonical transformation  of the bosonic coordinates 
  $(x^\mu, p_\mu,r,p_r,)\to (y^\mu,\pi_\mu,  R, p_R)$ with 
\be
R=\sqrt{\sum_{\nu=1}^8 x^\nu x^\nu}\;:=|\mathbf{x}|, \quad 
 p_R=\frac{\sum_{\mu=1}^8 p_\mu x^\mu}{|\mathbf{x}|}, \quad y^\mu= {r} \frac{x^\mu}{|\mathbf{x}|},\quad \pi_\mu=
\frac{ |\mathbf{x}|}{r}p_\mu +  \left(p_r-\frac{\sum_{\lambda =1}^8p_\lambda x^\lambda}{r}\right) \frac{x^\mu}{|\mathbf{x}| }
,
\ee
 we  bring the bosonic part of the Hamiltonian to the following form
 \be
\mathcal{H}_{bos}=\left.{\cal H}\right|_{{\chi }\to\;0}=\frac18 \sum_{\mu,\nu =1}^8 g^{\mu\nu } \pi_\mu\pi_\nu ,\quad {\rm with}\quad g^{\mu\nu}=\delta^{\mu\nu}+  \frac{3y^\mu y^\nu}{\sum_{\lambda=1}^8 y^\lambda y^\lambda},\quad
 L_{\mu\nu}=
\pi_\mu y^\nu-\pi_\nu y^\mu  .
\ee
 The Poisson brackets are defined by the relations
$
\{p_R, R\}=1$, $ \{\pi_\mu, y^\nu\}=\delta_\mu^\nu$
.
So, the variables $p_R, R$ do not appear in the Hamiltonian anymore, i.e. we deal with an eight-dimensional system.

From the above expression of Hamiltonian  we immediately get the metric of configuration space:
\be
(ds)^2=\sum_{\mu,\nu=1}^8 g_{\mu\nu}dy^\mu dy^\nu,\qquad g_{\mu\nu}=\delta_{\mu\nu}- \frac34\frac{y^\mu y^\nu}{ {\sum_{\lambda=1}^8y^\lambda y^\lambda}} ,\qquad g^{\mu\lambda}g_{\lambda\nu}=\delta^\mu_\nu.
\ee
It  is  just the induced metrics of the cone in the nine-dimensional pseudo-Euclidian space $\mathbf{R}^{8.1}$, 
\be
ds^2=-dy^0dy^0+ (d\mathbf{y}d\mathbf{y}) ,\qquad{\rm  where }\quad  4y_0= 3|\mathbf y|.
\ee
Performing  conformal transformation we can represent  this metrics to the conformal-flat form
\be
\tilde{y}^\mu=\frac{y^\mu}{\sqrt{|\mathbf{y}|}}\quad \Rightarrow\quad (ds)^2=|\mathbf{\tilde{y}}|^2 d\mathbf{\tilde{y}}d\mathbf{\tilde{y}}
\ee
Thus, the bosonic part of the Hamiltonian is  a free  particle on the eight-dimensional cone  embedded in    nine-dimensional pseudo-Euclidian space.

Let us also notice that the  fermionic part of Hamiltonian,  $ ({\sum_{\mu,\nu=1}^8 L_{\mu\nu }R_{\mu\nu}  })/( {\sum_{\lambda=1}^8y^\lambda y^\lambda}) $, can be interpreted as a spin-orbit coupling term.

\section{Conclusion}

In this paper,  we  give the Hamiltonian formulation to the $\cN=8$ supersymmetric mechanic constructed recently in \cite{IvFed}. The Hamiltonian description reveals some interesting  properties of the model hidden in the Lagrangian description. In particular, we demonstrate that
\begin{itemize}
	\item The system possesses the dynamical conformal $osp(8|2)$ supersymmetry,
	\item The Poincar\'e  supercharges  have a nice structure,  being just the product of $R$-symmetry generators to the fermions,
	\item The bosonic sector of the system describes a free particle on {\sl eight-dimensional   cone } in    nine-dimensional pseudo-Euclidian space . The fermionic part can be interpreted as  a spin-orbit coupling term.
\end{itemize}
This kind  of   a supersymmetric mechanics is  quite unusual  and has been   out of   focus in the  supersymmetric mechanics community. For  nontrivial examples of similar  systems we refer to \cite{bny} where the three-dimensional $\mathcal{N}=4$ supersymmetric mechanics with a spin-orbit coupling term were  constructed by the Hamiltonian reduction of the appropriate four-dimensional system. 
We have no doubts that  the earlier introduced $\mathcal{N}=4$ supersymmetric mechanics with "semi-dynamical" (isospin) variables  \cite{FIL-2009} can be interpreted in a similar way.
These systems can be directly obtained from the action \eqref{Hm}, \eqref{Ii} and   supercharges 
\eqref{sc}, \eqref{Theta}  in the limit $\left\{\chi^{i'A}, z_a^{jA} \right\} \rightarrow 0$ or
$\left\{\phi^i, B_a^{i'}\right\} \rightarrow 0$. In both cases we will end up  with $\cN=4$ supersymmetric mechanics with $osp(4|2)$ dynamical superconformal symmetry.
  
Another interesting issue is the reduction of the present system by the $SU(2)$  group action.
As a result of such reduction, we will arrive at the five-dimensional    system  specified by an additional interaction with $SU(2)$ Yang monopole (cf. \cite{toppan}). Also, it would be interesting to construct similar systems on the curved background as well as  to construct their  higher-dimensional ("multi-particle") analogs.

 It seems that  the presented system can be elegantly described in terms of octonions.
 The point  is that   two-dimensional $\mathcal{N}=2$ and four-dimensional  $\mathcal{N}=4$ supersymmetric systems with spin-orbit interaction term can be  surprisingly formulated in a unified way   in terms of complex and quaternionic variables, respectively.
  We are planning to   present these systems elsewhere, hopefully with their eight-dimensional octonionic $\mathcal{N}=8$  counterpart.

\acknowledgements
AN thanks Evgeny Ivanov and Sergey Fedoruk for  interest to this work and encouragement.
The work was supported  by the Armenian State Committee of Higher Education  and Science within the projects   21AG-1C062 (A.N., E.Kh.)  and 21AA-1C001 (E.Kh.).

\def\theequation{A.\arabic{equation}}
\setcounter{equation}{0}
\section*{Appendix A: Superconformal algebra $
\mathbf{Osp(8|2)}$}
Here we present the Poisson brackets relations  of the $\mathcal{N}=8$ conformal superalgebra $osp(8|2)$. The $osp(8|2)$ superalgebra contains 31 bosonic generators forming the $sl(2) \times so(8)$ algebra and 16 fermionic generators - 8 generators $\left\{ Q^i, {\bar Q}_i, Q^{i' A}\right\}$ forming the $d=1, \;\mathcal{N}=8$  Poincar\'e superalgebra and 8 generators $\left\{ S^i, {\bar S}_i, S^{i' A}\right\}$ of conformal supersymmetry. 

The $sl(2)$ generators span the $d=1$ conformal algebra, while $so(8)$ generators form the 
$R$-symmetry algebra. We identify the $sl(2)$ generators ${\cal H}, D, K$ with the time-translation, dilatation and conformal boost, respectively. We divided the $so(8)$ generators
into four mutually commuting $su(2)$ algebras $\left\{J_3,J, {\bar J}\right\},  X^{i'j'}, W^{ij},Y^{AB}$ and the generators $V^{ij'A}$ and ${\bar V}^{ij'A}$ belonging to the coset $so(8)/su(2)^4 $. The commutators of the $sl(2) \times so(8)$ algebra in our setup look as follows:
\subsection{Bosonic sector}
\begin{itemize}
	\item{The conformal $sl(2)$} algebra contains the generators ${\cal H}, D, K$ which obey the following Poisson brackets
	\be
	\left\{ D, {\cal H}\right\} = -{\cal H}, \;	\left\{ D, K\right\} =K, \;	\left\{ {\cal H},K\right\} = 2 D.
	\label{ca}\ee
	The generators ${\cal H}, D, K$ are Hermitian   and have trivial Poisson brackets with the generators of the $so(8)$ algebra.
	
	\item{First $su(2)$}
	\be  
\left\{ J, {\bar J} \right\} = 2\imath J_3, \;\left\{ J_3, J\right\}=   \imath J, \;\left\{ J_3, {\bar J} \right\}=  -\imath {\bar J}.
	\label{so81}\ee
	The generators $J, {\bar J}, J_3$ have trivial Poisson brackets with the generators $ X^{i'j'}, W^{ij},Y^{AB}$, while
	the brackets with  $V^{ij'A}$ and ${\bar V}^{ij'A}$ read
	\be 
	\left\{ J, {\bar V}^{ij'A} \right\} =  V^{ij'A},\; 
	\left\{ {\bar J},  V^{ij'A} \right\} =-  {\bar V}^{ij'A},\; 
	\left\{J_3 , V^{ij'A} \right\} = \frac{\imath}{2} V^{ij'A},\; \left\{J_3 , {\bar V}^{ij'A} \right\} = -\frac{\imath}{2} {\bar V}^{ij'A}.
	\ee
	\item{Second $su(2)$}
	\be  
	\left\{ X^{i'j'}, X^{k'l'}\right\} = -\epsilon^{i'k'} X^{j'l'}- \epsilon^{j'l'} X^{i'k'} .
	\ee
	The generators $X^{i'j'}$  have trivial Poisson brackets  with the generators $ W^{ij},Y^{AB}$, while the brackets with  $V^{ij' A}$ and ${\bar V}^{i j' A}$ read
	\be 
	\left\{X^{i'j'},V^{k m' A} \right\} = -\frac{1}{2} \epsilon^{i' m'} V^{k j' A} -\frac{1}{2} \epsilon^{j' m'} V^{k i' A}, \quad 
	\left\{X^{i'j'},{\bar V}^{k m' A} \right\} = -\frac{1}{2} \epsilon^{i' m'} {\bar V}^{k j' A} -\frac{1}{2} \epsilon^{j' m'} {\bar V}^{k i' A}.
	\ee
	\item{Third $su(2)$}
	\be  
	\left\{ W^{ij}, W^{kl}\right\} = -\epsilon^{ik} W^{jl}- \epsilon^{jl} W^{ik} .
	\ee
	The generators $W^{ij}$  have trivial Poisson brackets  with the generators $Y^{AB}$, while
	the brackets with  $V^{i j' A}$ and ${\bar V}^{i j' A}$ read
	\be 
	\left\{W^{ij},V^{k m' A} \right\} = -\frac{1}{2} \epsilon^{ik} V^{j m' A} -\frac{1}{2} \epsilon^{jk} V^{i m' A}, \quad 	 
	\left\{W^{ij},{\bar V}^{k m' A} \right\} = -\frac{1}{2} \epsilon^{ik} {\bar V}^{j m' A} -\frac{1}{2} \epsilon^{jk} {\bar V}^{i m' A}.
	\ee
	\item{Fourth $su(2)$}
	\be  
	\left\{ Y^{AB}, Y^{CD}\right\} = -\epsilon^{AC} Y^{BD}- \epsilon^{BD} Y^{AC} .
	\ee
	The Poisson brackets with the generators  $V^{ij'A}$ and ${\bar V}^{ij'A}$ read
	\be
	\left\{ Y^{AB},V^{i j' C} \right\} = -\frac{1}{2} \epsilon^{AC} V^{i j' B} -\frac{1}{2} \epsilon^{BC} V^{i j' A}, \quad 	 
	\left\{Y^{AB},{\bar V}^{i j' C} \right\} = -\frac{1}{2} \epsilon^{AC} {\bar V}^{i j' B} -\frac{1}{2} \epsilon^{BC} {\bar V}^{i j' A}.
	\ee
	\item Finally, the generators $V^{i j' A}$ and ${\bar V}^{i' j' A}$, from the coset
	$so(8)/su(2)^4$, obey the following brackets:
	\bea
	&& \left\{ V^{i j' A} ,V^{k l' B} \right\} = -2 \epsilon^{ik}\epsilon^{j'l'} \epsilon^{AB} J, \;
	\left\{ {\bar V}^{i j' A} ,{\bar V}^{k l' B} \right\} = -2 \epsilon^{ik}\epsilon^{j'l'} \epsilon^{AB} {\bar J}, \nonumber \\
	&& \left\{ V^{i j' A} ,{\bar V}^{k l' B} \right\} = 2 \epsilon^{j'l'} \epsilon^{AB} W^{ik} +
	2 \epsilon^{ik}\epsilon^{j'l'} Y^{AB}+ 2	\epsilon^{ik} \epsilon^{AB} X^{j'l'}- 2 \imath \epsilon^{ik} \epsilon^{j'l'} \epsilon^{AB} J_3. 
	\label{so8last}\eea
\end{itemize}

The Poisson brackets between the fermionic generators look like 
\subsection{Fermionic sector}
\begin{itemize}
	\item the $d=1,\;\mathcal{N}=8$  Poincar\'e superalgebra
	\bea
	&& \left\{Q^i , {\bar Q}^j \right\} = -2 \imath\epsilon^{ij} {\cal H}, \quad \left\{Q^i,Q^j\right\} =\left\{{\bar Q}^i, {\bar Q}^j\right\} = 	\left\{Q^{i}, {\cal H}\right\} =	\left\{{\bar Q}^{i}, {\cal H}\right\} =0, \nonumber\\
	&&  \left\{Q^{i' A}, Q^{j' B}\right\} = 4   \imath \epsilon^{i'j'} \epsilon^{AB} \, {\cal H}, \quad
	\left\{Q^{i' A}, Q^j\right\} = \left\{Q^{i' A}, {\bar Q}^j\right\}  =	\left\{Q^{i' A}, {\cal H}\right\} =0.
	\label{n8}\eea
	\item  Superconformal 
	\bea
	&& \left\{S^i , {\bar S}^j \right\} = -2 \imath \epsilon^{ij} K, \quad \left\{S^i, S^j\right\} =\left\{{\bar S}^i, {\bar S}^j\right\} = 	\left\{S^{i}, K \right\} =	\left\{{\bar S}^{i}, K \right\} =0, \nonumber \\
	&&  \left\{S^{i' A}, S^{j' B}\right\} = 4 \imath\epsilon^{i'j'} \epsilon^{AB}   K, \quad
	\left\{S^{i' A}, S^j\right\} = \left\{S^{i' A}, {\bar S}^j\right\}  =
	\left\{S^{i' A}, K\right\} =0.
	\eea
	\item Cross commutators 
	\bea
	&& \left\{ Q^i, S^j \right\} = \imath\epsilon^{ij} J ,\; \left\{ {\bar Q}^i, {\bar S}^j \right\} = \imath \epsilon^{ij} {\bar J} , \nonumber \\
	&& \left\{ Q^i, {\bar S}^j \right\} = -2 \imath\epsilon^{ij} D - \imath W^{ij} - \epsilon^{ij} J_3, \;  \left\{ {\bar Q}^i, S^j \right\} = 2 \imath\epsilon^{ij} D + \imath W^{ji} - \epsilon^{ij} J_3, \nonumber \\
	&& \left\{ Q^i, S^{j' A} \right\} = \imath V^{i j' A} \;,  \left\{{\bar Q}^{i}, S^{j' A}  \right\} = \imath {\bar V}^{i j'  A} , \nonumber \\
	&& \left\{  Q^{i' A}, S^j, \right\} = -\imath  V^{j i' A},  \;  \left\{ Q^{i' A}, {\bar S}^j \right\} = -{\bar V}^{j i' A}, \nonumber \\
	&&\left\{ Q^{i' A}, S^{j' B}\right\} = 4\imath \epsilon^{i'j'} \epsilon^{AB} D +2 \imath \epsilon^{AB}X^{i'j'}+
	2 \imath \epsilon^{i'j'} Y^{AB} .
	\eea
\end{itemize}
Finally, the cross 	boson-fermion brackets have the following structure.
\subsection{ Boson-Fermion Poisson brackets}
\begin{itemize}
	\item $sl(2) \times \mbox{Supercharges}$
	\bea
	&&\left\{{\cal H},Q^i\right\} =\left\{{\cal H},{\bar Q}^i\right\} =\left\{{\cal H},Q^{i' A}\right\} = 0,\quad
	\left\{{\cal H},S^i\right\} =Q^i, \; \left\{{\cal H},{\bar S}^i\right\} ={\bar Q}^i,  \; \left\{{\cal H},S^{i' A}\right\} = Q^{i' A} , \nonumber \\
	&&\left\{D,Q^i\right\} =-\frac{1}{2} Q^i, \;\left\{D,{\bar Q}^i\right\} =-\frac{1}{2} {\bar Q}^i, \;\left\{D,Q^{i' A}\right\} =-\frac{1}{2} Q^{i' A}, \nonumber \\
	&& \left\{D,S^i\right\} =\frac{1}{2} S^i, \;\left\{D,{\bar S}^i\right\} =\frac{1}{2} {\bar S}^i, \;\left\{D,S^{i' A}\right\} =\frac{1}{2} S^{i' A},  \nonumber\\
	&&\left\{K,S^i\right\} =\left\{K,{\bar S}^i\right\} =\left\{K,S^{i' A}\right\} = 0,\quad
	\left\{K,Q^i\right\} =-S^i, \; \left\{K,{\bar Q}^i\right\} =-{\bar S}^i,  \; \left\{K,Q^{i' A}\right\} = -S^{i' A}.
	\eea
	\item First $su(2) \times \mbox{Supercharges}$
	\bea
	&& \left\{J,Q^i\right\}= \left\{J,S^i\right\}= \left\{J,Q^{i' A}\right\}= \left\{J,S^{i' A}\right\}=0,\quad   \left\{J,{\bar Q}^i\right\}=  Q^i,\;  \left\{J,{\bar S}^i\right\}=  S^i, \nonumber \\
	&& \left\{J_3,Q^i\right\}=\frac{\imath}{2}Q^i,\; \left\{J_3,{\bar Q}^i\right\} =-\frac{\imath}{2}{\bar Q}^i, \;
 \left\{J_3,S^i\right\}=  \frac{\imath}{2}S^i, \; \left\{J_3,{\bar S}^i\right\} =-\frac{\imath}{2}{\bar S}^i,\quad
	 \left\{J_3,Q^{i' A}\right\}= \left\{J_3,S^{i' A}\right\}=0, \nonumber \\
	&& \left\{{\bar J},{\bar Q}^i\right\}= \left\{{\bar J},{\bar S}^i\right\}= \left\{{\bar J},Q^{i' A}\right\}= \left\{{\bar J},S^{i' A}\right\}=0,\quad   \left\{{\bar J}, Q^i\right\}= - {\bar Q}^i,\;  \left\{{\bar J}, S^i\right\}= - {\bar S}^i, 
	\eea
	\item Second $su(2) \times \mbox{Supercharges}$
	\bea
	&&\left\{X^{i'j'},Q^k\right\}=\left\{X^{i'j'},{\bar Q}^k\right\}=\left\{X^{i'j'},S^k\right\}=\left\{X^{i'j'},{\bar S}^k\right\}=0, \nonumber\\
	&&\left\{X^{i'j'},Q^{k' A}\right\}=-\frac{1}{2}\left( \epsilon^{i'k'}Q^{j' A}+ \epsilon^{j'k'}Q^{i' A}\right),\;
	\left\{X^{i'j'},S^{k' A}\right\}=-\frac{1}{2}\left( \epsilon^{i'k'}S^{j' A}+ \epsilon^{j'k'}S^{i' A}\right).
	\eea
	\item Third $su(2) \times \mbox{Supercharges}$
	\bea
	&&\left\{W^{ij},Q^k\right\}=-\frac{1}{2}\left(\epsilon^{ik}Q^j+\epsilon^{jk}Q^i\right),\;
	\left\{W^{ij},{\bar Q}^k\right\}=-\frac{1}{2}\left(\epsilon^{ik}{\bar Q}^j+\epsilon^{jk}{\bar Q}^i\right),\quad 	\left\{W^{ij},Q^{k' A}\right\} =0, \nonumber \\
	&&\left\{W^{ij},S^k\right\}=-\frac{1}{2}\left(\epsilon^{ik} S^j+\epsilon^{jk} S^i\right),\;
	\left\{W^{ij},{\bar S}^k\right\}=-\frac{1}{2}\left(\epsilon^{ik}{\bar S}^j+\epsilon^{jk}{\bar S}^i\right),\quad 	\left\{W^{ij},S^{k' A}\right\} =0. 
	\eea
	\item Fourth $su(2) \times \mbox{Supercharges}$
	\bea
	&& \left\{Y^{AB}, Q^i\right\}= \left\{Y^{AB}, {\bar Q}^i\right\}= \left\{Y^{AB}, S^i\right\}= \left\{Y^{AB},{\bar S}^i\right\}= 0, \nonumber \\
	&& \left\{ Y^{AB},Q^{i' C}\right\} = -\frac{1}{2} \left( \epsilon^{AC}Q^{i' B}+\epsilon^{BC}Q^{i' A}\right), \;
	\left\{ Y^{AB},S^{i' C}\right\} = -\frac{1}{2} \left( \epsilon^{AC}S^{i' B}+\epsilon^{BC}S^{i' A}\right).
	\eea
	\item The brackets of $V^{i j' A}$ and ${\bar V}^{i j' A}$ with Supercharges
	\bea
	&&\left\{V^{ij'A},Q^k \right\} =0, \; \left\{V^{ij'A},{\bar Q}^k \right\} =\epsilon^{ik}Q^{j' A},\;
	\left\{{\bar V}^{ij'A},{\bar Q}^k \right\} =0, \; \left\{{\bar V}^{ij'A}, Q^k \right\} =-\epsilon^{ik}Q^{j' A}, \nonumber \\
	&&\left\{V^{ij'A},S^k \right\} =0, \; \left\{V^{ij'A},{\bar S}^k \right\} =\epsilon^{ik} S^{j' A},\;
	\left\{{\bar V}^{ij'A},{\bar S}^k \right\} =0, \; \left\{{\bar V}^{ij'A}, S^k \right\} =-\epsilon^{ik} S^{j' A}, \nonumber \\
	&&\left\{V^{ij'A}, Q^{k'B}\right\} = 2 \epsilon^{j'k'} \epsilon^{AB} Q^i, \;
	\left\{{\bar V}^{ij'A}, Q^{k'B}\right\} = 2 \epsilon^{j'k'} \epsilon^{AB} {\bar Q}^i, \nonumber \\
	&&\left\{V^{ijA}, S^{kB}\right\} = 2 \epsilon^{jk} \epsilon^{AB} S^i, \;
	\left\{{\bar V}^{ijA}, S^{kB}\right\} = 2 \epsilon^{jk} \epsilon^{AB} {\bar S}^i .
	\eea
\end{itemize}

\end{document}